\documentclass[conference]{IEEEtran}
\IEEEoverridecommandlockouts

\usepackage{cite}
\usepackage{graphicx}
\usepackage{float}
\usepackage{amsmath,amssymb,amsfonts}
\usepackage{algorithmic}
\usepackage{graphicx}
\usepackage{soul}
\usepackage{textcomp, comment, soul}
\usepackage[inline]{enumitem} 
\usepackage[mathscr]{euscript}
\usepackage{mathrsfs}
\usepackage{ulem}

\usepackage{array, makecell, booktabs}
\usepackage[table,xcdraw]{xcolor}
\usepackage{tikz}

\usepackage[colorlinks=true, allcolors=blue]{hyperref}

\def\BibTeX{{\rm B\kern-.05em{\sc i\kern-.025em b}\kern-.08em
    T\kern-.1667em\lower.7ex\hbox{E}\kern-.125emX}}

\begin{document}

\title{Hardware-Efficient EMG Decoding for Next-Generation Hand Prostheses\\
\vspace*{-0.25\baselineskip}
}
\author{
\IEEEauthorblockN{
Mohammad Kalbasi$^{1,2}$, MohammadAli Shaeri$^1$, Vincent Alexandre Mendez$^{3}$, Solaiman Shokur$^{3}$,\\ Silvestro Micera$^{3}$, Mahsa Shoaran$^1$
}
\IEEEauthorblockA{
$^1$Institute of Electrical and Micro Engineering and Neuro-X Institute, EPFL, Geneva, Switzerland\\
$^2$Department of Electrical Engineering, Sharif University of Technology, Tehran, Iran \\
$^3$Bertarelli Foundation Chair in Translational Neuroengineering, Neuro-X Institute, EPFL, Geneva, Switzerland\\
Email: \{mohammad.kalbasi, mohammad.shaeri, mahsa.shoaran\}@epfl.ch
}
\vspace*{-1.75\baselineskip}
}
\maketitle
\begin{abstract}
Advancements in neural engineering have enabled the development of Robotic Prosthetic Hands (RPHs) aimed at restoring hand functionality.
Current commercial RPHs offer limited control through basic on/off commands.
Recent progresses in machine learning enable 
finger movement decoding with higher degrees of freedom,
yet the high computational complexity of such models limits their application in portable devices. Future RPH designs must balance portability, low power consumption, and high decoding accuracy to be practical for individuals with disabilities.
To this end, we introduce a novel attractor-based neural network to realize on-chip movement decoding for next-generation portable RPHs.
The proposed architecture comprises an encoder, an attention layer, an attractor network, and a refinement regressor.
We tested our model on four healthy subjects and achieved a decoding accuracy of 80.6\textpm3.3\%.
Our proposed model is over 120 and 50 times more compact compared to state-of-the-art LSTM and CNN models, respectively, with comparable (or superior) decoding accuracy.
Therefore, it exhibits minimal hardware complexity and can be effectively integrated as a System-on-Chip. 



\end{abstract}


\section{Introduction}
Losing a hand is a devastating experience, profoundly affecting both a human's physical and emotional well-being\cite{b1,grob2008psychological}. 
Robotic Prosthetic Hands (RPHs) enhance amputees' quality of life by restoring hand functionality.
Such devices should mimic natural movements to induce sensations of the human hand \cite{b2,b3}. 
Hand prostheses first record Electromyographic (EMG) signals from a subject's forearm, which consists of electrical activities generated by muscle contractions.
In a classic example, the EMG signal recorded from two surface EMG electrodes (sEMG) was converted into a binary On/Off command for a robotic hand using a simple thresholding technique, enabling control with one degree of freedom \cite{b4,b5,mendez2021current}. 
Modern High-Density sEMG (HD-sEMG) electrodes are capable of recording from 128 channels and offer the potential to decode fine and dexterous movements from muscle activities \cite{malevsevic2021database}.
Conventional EMG decoders have typically been developed for classifying different hand gestures \cite{b6,b7,zabihi2023trahgr,esposito2020piezoresistive}.
In \cite{kim2008emg}, a combined model incorporating kNN and Bayes classifiers was employed to decode four classes of hand gestures. 
Recently, a vision transformer model was utilized to classify as many as 65 hand gestures \cite{b8}. 
While such models achieve high accuracy in movement classification tasks, 
they lack the ability to decode a continuous movement trajectory,  hindering the naturalness of prosthetic hand movements.\\

To overcome this limitation, other studies have introduced AI solutions to continuously predict finger angles over time \cite{b9,b10,smith2008continus,ngeo2014continuous,pan2014continuous}.
In \cite{stapornchaisit2019finger}, Independent Component Analysis (ICA) was used to isolate muscle activities associated with finger movements while a linear regressor predicted finger extension and flexion from forearm EMG. 
In another study, a Convolutional Neural Network (CNN) was utilized to simultaneously decode finger movements \cite{b11}.
Given the proficiency of Recurrent Neural Networks (RNNs) in capturing intricate patterns,  an encoder-decoder algorithm incorporating a Gated Recurrent Unit (GRU) with an attention mechanism was used to predict 14 finger joint angles \cite{b14}. 
Moreover, a Long Short-Term Memory (LSTM) model accompanying manifold learning and attention mechanism demonstrated promising results in decoding 22 joint angles \cite{b15}.\\
In general, deep neural networks such as CNN and LSTM are recognized for their effectiveness in EMG decoding \cite{b11, ameri2019regression, azhiri2021realtime}.
These methods are compute-intensive software solutions, requiring an external computer to process and decode EMG signals.
The next generation of RPHs will offer seamless, portable prosthetic solutions that are user-friendly for individuals coping with hand loss, facilitating their daily activities.
The evolution towards next-generation portable RPHs necessitates efficient integration with AI to directly translate EMG signals into movement commands, eliminating the need for an external computer. This approach follows the recent advancements in the intelligent neural interface domain with embedded AI \cite{shoaranenergy, SSCM, CL, NT}.
Hence, model size and complexity must be prioritized in addition to decoding accuracy to ensure hardware efficiency \cite{b17, Yoo, CICC, RT}. 
To address these challenges, we propose a Dual Predictive Attractor-Refinement Strategy (DPARS) model 
for decoding continuous finger angles from EMG.
Our proposed model achieves a significantly lower complexity by reducing the number of computations while maintaining accuracy comparable to state-of-the-art models such as CNN and LSTM.
\begin{figure*}
\vspace*{-.2\baselineskip}
\centering
\includegraphics[width=.92\linewidth]{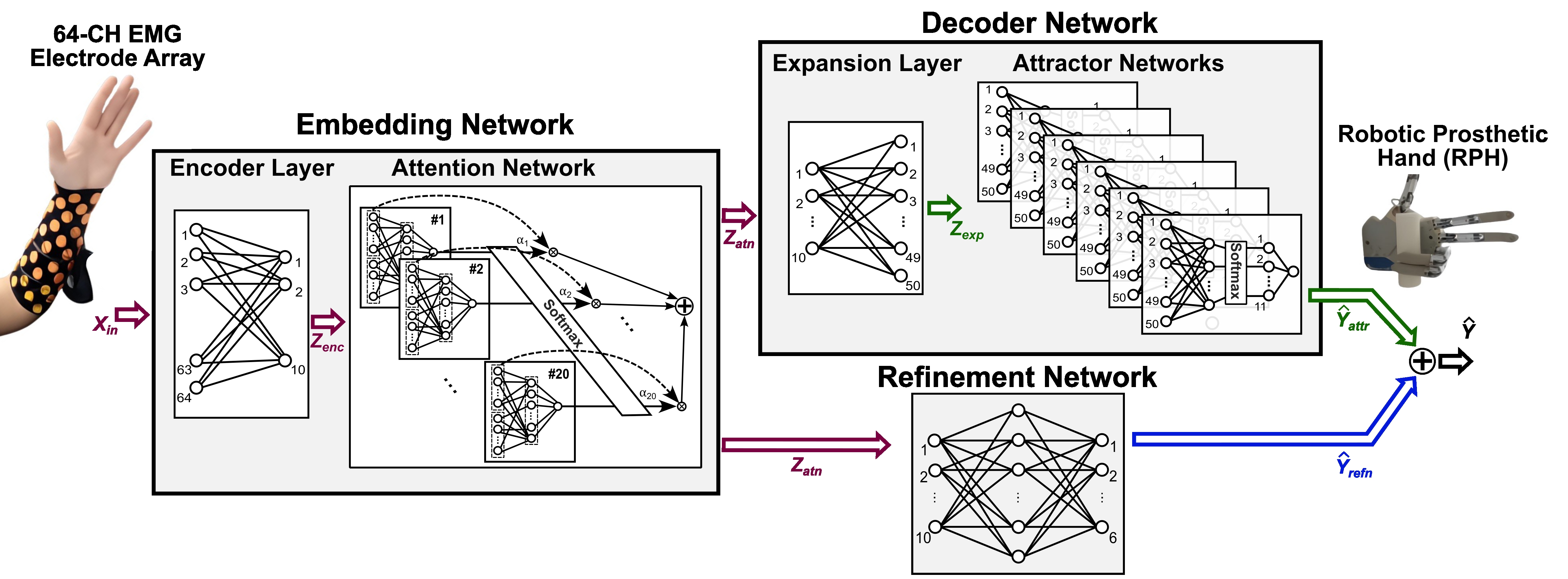}
\vspace*{-.8\baselineskip}
\caption{\label{fig:model}Dual Predictive Attractor-Refinement Strategy (DPARS) for real-time EMG decoding. 
The model includes an encoder layer for dimensionality reduction, followed by an attention network that highlights the most important sample in a data stream. Then, the attractor network generates a coarse prediction ($\hat{Y}_{\mathrm{attr}}$), while the refinement network refines the initial prediction by compensating for errors ($\hat{Y}_{\mathrm{refn}}$). The decoded finger angle ($\hat{Y}$) will be generated by summing the coarse and the refinement signal.}
\vspace*{-1\baselineskip}
\end{figure*}
\section{Model Architecture}\label{section:Model Architecture}
The dual predictive algorithm is designed to simultaneously estimate coarse movements (using an attractor-based decoder) and fine motions (using an MLP regressor) to restore natural, human-like control of RPHs.
As depicted in Fig. \ref{fig:model}, the proposed model predicts finger angles through the following steps:
\begin{enumerate*}[label=(\roman*),nosep]
\itemsep -0.5\parsep
\item An encoder network transforms the data representation from input space into a compact and informative embedding space.
\item An attention network enables the model to focus on relevant information in the temporal sequence of data. 
\item A novel attractor network computes attractor state probabilities and generates a coarse estimation of the finger angles. 
\item A refinement network enhances the overall performance by assessing the initial prediction's errors. 
\end{enumerate*}
This two-level prediction mechanism, involving an attractor network and a refinement network, effectively ensures a robust and accurate prediction of finger angles.

\subsection{Encoder Layer}
Dimensionality reduction techniques, such as Independent Component Analysis (ICA) or Principal Component Analysis (PCA), have been demonstrated to be essential in EMG decoding  \cite{stapornchaisit2019finger, dela2022emg, caesarendra2017emg}. 
Such techniques compress data by extracting informative features \cite{TransformReview2022RBME}.
They can enhance prediction accuracy in addition to lowering the computational burden of following blocks in the decoding pipeline. 
Here, we designed a Single-Layer Perceptron (SLP) that functions as an encoder, efficiently mapping the EMG data into a compact representation. The SLP performs a linear transformation through matrix multiplication as follows:
\begin{equation}
\label{eq:Linear_transformation}
{Z}_{\mathrm{enc}}[i] = W_{\mathrm{enc}} \times {X}_{\mathrm{in}}[i]
\end{equation}
where $W_{\mathrm{enc}}$ and ${X}_{\mathrm{in}}[i]$ are the encoder weight matrix and the input at timestep $i$, respectively. Encoder weights are computed through an end-to-end training procedure using a global objective function (more details in Section \ref{section:Model Architecture}.C). 
Thanks to dimensionality reduction, 
the decoder solely processes low-dimensional encoded data.
Thus, the number of computations is significantly reduced, resulting in a lower hardware cost.

\subsection{Attention Network}
The attention network learns patterns 
and assigns distinct weights to elements in a data sequence, emphasizing the most significant ones \cite{niu2021review}. 
In this architecture, the attention mechanism is implemented using Multi-Layer Perceptrons (MLPs) with two layers that process a data stream comprising the current encoded data sample and the 19 preceding samples. 
The attention network consists of 20 sub-networks, each taking the two encoded data vectors at current time 
(${Z}_{\mathrm{enc}}[i]$) and $j^{th}$ previous sample (${Z}_{\mathrm{enc}}[i\!\!-\!\!j]$) and calculating $S_{\mathrm{atn}}$ as:
\vspace*{-.25\baselineskip}
\begin{equation}
\label{eq:attention_layer}
S_{\mathrm{atn}}[i,j] = W_{\mathrm{atn}, 2} \times \mathrm{tanh}\left(W_{\mathrm{atn}, 1}
\begin{bmatrix}
{Z}_{\mathrm{enc}}[i\!-\!j] \\
{Z}_{\mathrm{enc}}[i]
\end{bmatrix}\right)
\vspace*{-.25\baselineskip}
\end{equation} 
where $W_{\mathrm{atn}}$ 
indicates  network weights and $S_{\mathrm{atn}}[i\!\!-\!\!j]$ shows the attention scores for $j^{th}$ previous sample in the sequence.
Using a softmax function, normalized attention coefficients
($[\alpha[i,0], ..., \alpha[i,19]] =\mathrm{Softmax}\left(S_{\mathrm{atn}}[i,0], ...,S_{\mathrm{atn}}[i,19]\right)$) 
are extracted, representing the importance of each timestep. 
Then, a context vector as a weighted sum of attention coefficients and input sequence is determined:
\vspace*{-.4\baselineskip}
\begin{equation}
\label{eq:zat}
{Z}_{\mathrm{atn}}[i] = \sum_{j = 0}^{19}\alpha[i,j]\times {Z}_{\mathrm{enc}}[i-j].
\vspace*{-.5\baselineskip}
\end{equation} 
\noindent The context vector conveys a more condensed and information-rich representation of the encoded data sequence.
It is worth mentioning that the attention network compresses the data stream in the time domain while the encoder reduces data dimensionally across input channels. 
The low-complexity MLP model implemented in this structure comprises a single hidden layer and an output layer. 
Moreover, the model is constrained to employ an identical network (i.e., with the same structure and weights) throughout training.
Therefore, it can be implemented as a single MLP network and reused at different timesteps, resulting in a 20$\times$ reduction in hardware cost compared to the conventional design.

\subsection{Attractor Network}
Fingers naturally exhibit a tendency to maintain specific positions between flexed and extended states despite their free movement.
This resembles attractors, a concept used in mathematics to describe the states  
that a dynamical system tends to end up in \cite{b19,b20}.
\
Inspired by this concept, we introduce a novel attractor-based network that extracts the most probable states (attractors) and predicts the output based on their probability distributions. 
To address complexity and accuracy concerns, we limit the search domain for attractor extraction to eleven discrete states, encompassing a range from fully flexed (90$^\circ$) to fully extended (180$^\circ$).
\
Figure \ref{fig:model} illustrates the attractor networks within a decoder network positioned after an expansion layer.
The expansion layer enhances the model's capability to extract more intricate patterns by mapping the context vector to a higher-level feature space. 
Then, the attractor network utilizes two-layer MLPs to assess the likelihood of each individual state in the attractor set (\(\mathscr{K}\)) for each finger.
It generates an initial estimate of a finger angle by computing a sum of attractor values weighted by their associated likelihoods, expressed as:
\vspace*{-.1\baselineskip}
\begin{equation}
\label{eq:yhat}
\hat{Y}_{\mathrm{attr}}^{c} = \sum_{k \in \mathscr{K} }k\times Pr({Y_{q}^{c}} = k )
\vspace*{-.1\baselineskip}
\end{equation}
where $Y_q^c$ is a random variable representing the likelihood of each discrete attractor state for the corresponding finger $c$.
The probability distribution of $Y_q^c$ is generated by the attractor network, quantifying the likelihoods for each state and determining the probabilistic prediction of finger angles.

The attractor-based decoder is trained along with the embedding network to predict outputs by leveraging the most probable states. 
The model is trained using an objective function aimed at minimizing the $l_1$ loss and identifying the attractors. 
The objective function is defined as:
\vspace*{-.5\baselineskip}
\begin{equation}
\label{eq:loss_func}
G(\theta) = 
\underset{X,Y}{\mathbb{E}}\Big[\ \lVert\hat{Y}(X;\theta) - Y\rVert_1 + \lambda \sum_{c = 1}^{6}H\left(Y_{q}^{c}(X;\theta)\right)\ \Big]
\vspace*{-.5\baselineskip}
\end{equation}
where $Y$ and $\hat{Y}$ are the real and predicted finger angles, respectively. 
$\theta$ denotes the parameter set of the model and $H$ is the entropy operator. 
The entropy term has a key role in extracting attractors by assigning only a few highly probable states to the attractor set. 

\subsection{refinement Network}
The decoder network provides a coarse yet accurate initial estimation based on attractors.
In parallel, a refinement network is incorporated into the model to improve the naturalness of RPH movements by capturing fine details.
In this design, a two-layer MLP regressor is used to predict complementary movements for each finger angle. 
By adding this signal to the initial prediction, the DPARS output is determined as follows:
\vspace*{-.4\baselineskip}
\begin{equation}
\label{eq:yhat_2}
\hat{Y} = \hat{Y}_{\mathrm{attr}} + \hat{Y}_{\mathrm{refn}}
\vspace*{-.4\baselineskip}
\end{equation}
where ${Y}_{\mathrm{attr}}$ and ${Y}_{\mathrm{refn}}$ denote the output of the attractor network and refinement network, respectively.
By summing these two signals, 
the proposed model effectively decodes the coarse and fine activities in intricate movements. 
This dual predictive strategy enhances the prediction accuracy, robustness, and naturalness of the movement decoding.

\begin{figure}
\centering
\includegraphics[width=0.7\linewidth]{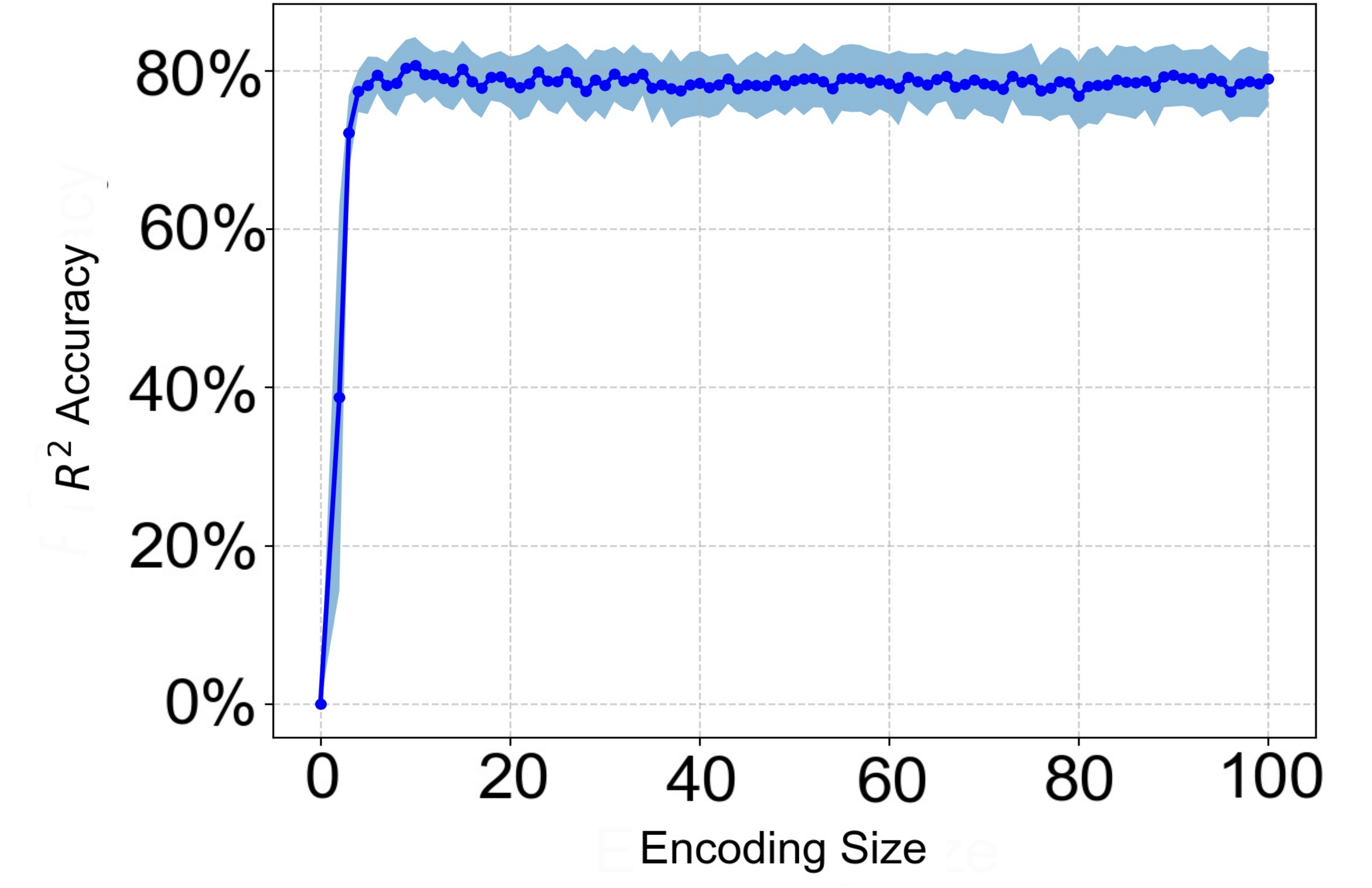}
\vspace*{-.9\baselineskip}
\caption{\label{fig:encoder}Impact of Encoding Size on Decoding Accuracy: The blue line represents the mean $R^2$ accuracy against the encoding size, and the shaded area indicates the variance. 
}
\vspace*{-1\baselineskip}
\end{figure}

\section{Results}
In this study, we recorded EMG and finger movements from four healthy subjects, ethically approved by the Cantonal Ethical Committee of Vaud, while all participants provided informed consent for participating in the study. Each subject was tasked with replicating various finger movements and grasps, as demonstrated in the pre-captured videos provided by our research team. These movements involved flexing, extending, or maintaining a middle position (rest) of the fingers.
The video sequences were repeated six times. Participants were instructed to mimic the finger movements displayed in the videos. The first four repetitions were dedicated to the training dataset.
The fifth and sixth repetitions were separately allocated to the validation and test datasets, ensuring robust evaluation of the model's predictive performance.

The EMG data was acquired using a Medium Density EMG (MD-EMG) system with 64 monopolar channels and dry electrodes \cite{mendez2023improving}. The signals were captured at a frequency of 2400 Hz and filtered with a bandpass filter within a range of 5 to 500 Hz, along with a 50 Hz notch filter to remove the power line interference. Next, the envelope of the EMG signal was extracted by applying a low-pass filter to the absolute value of the signal and then downsampled at a rate of 100 Hz ($T_s$=10ms).
The input data stream is then segmented into 200 ms data blocks with 190 ms overlap between windows, which is equal to 20 data samples per block for predicting each output.
The proposed DPARS algorithm is implemented using the PyTorch toolkit in Python. 
All blocks in DPARS are designed based on feedforward fully connected MLP architecture to optimize computational efficiency and minimize model complexity.
The model parameters are learned by minimizing the entropy-regularized loss function described in Eq. (\ref{eq:loss_func}) using the Stochastic Gradient Descent (SGD) optimizer with a batch size of 64 for 100 epochs. During model training, the parameters yielding the lowest validation loss are ultimately used as the final configuration of the model. 
The performance is assessed with the normalized error measured by the $R^2$ metric on the test sets between predicted and real finger angles.

\subsection{Computational Efficiency of the Model} 
The encoder network transforms the 64-ch EMG data into a more compact and informative representation. 
Figure \ref{fig:encoder} illustrates the prediction accuracy against the encoding size.
DPARS achieves its optimal performance with an encoding size of ten, resulting in a significant reduction of 6.4$\times$ in dimensionality. Then, the attention network processes encoded data for 20 consecutive timesteps (i.e., with 20$\times$10 dimensions) and generates a context vector (i.e., with a dimension of 10), yielding an additional 20$\times$ dimensionality reduction. 
Thanks to reducing encoding size and data dimensionality, the number of Multiply–Accumulate (MAC) operations and the required memory are decreased by a factor of 128 (6.4$\times$20) in the subsequent layers.

\begin{figure}
\centering
\includegraphics[width=0.9\linewidth]{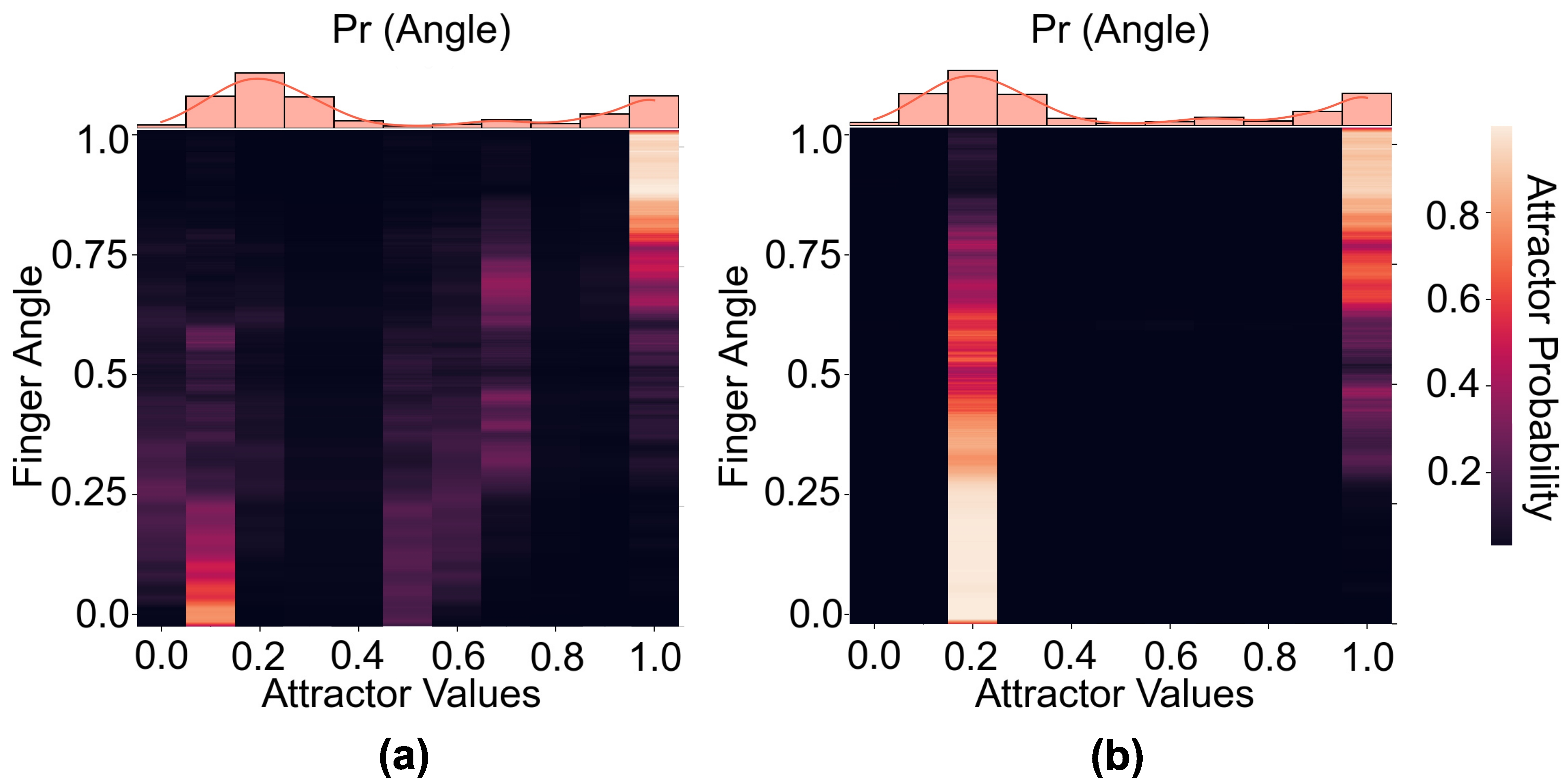}
\vspace*{-0.75\baselineskip}
\caption{\label{fig:attractor}Probability distributions generated by the attractor network. (a) Probability distribution spread among all possible states through minimizing decoder loss only. (b) The network extracts attractors. The probability of non-attractor states equals zero when the entropy regularization term is included in the objective function, leading to \textgreater4$\times$ less number of computations. 
\vspace*{-0.5\baselineskip}
}
\end{figure}

The attractor network induces sparsity in data representation by penalizing high entropy values. 
As a result, the probabilities are only assigned to a specific attractor, and the output is a weighted sum of the attractor states.
This has the potential to reduce the hardware cost and the number of MAC operations by \textgreater4$\times$.
Figure \ref{fig:attractor}(a) illustrates the state probabilities estimated without entropy regularization.
Incorporating the entropy term into the objective function (eq. (\ref{eq:loss_func})), the data distribution is notably concentrated on two distinct states (attractors) as shown in Fig. \ref{fig:attractor}(b). 
The attractor network essentially captures attractors 
and generates a relatively coarse prediction.
It allows the refinement network to focus on fine details and refine the coarse prediction.
This dual predictive strategy incorporating entropy regularization boosts decoding accuracy by more than 3\%, as shown in Table \ref{table: result}.

\begin{table}[b]
\vspace*{-1\baselineskip}
\begin{centering}
\caption{\label{table: result}Performance comparison of the proposed model and state-of-the-art models}
\vspace*{-.75\baselineskip}
\centering

\resizebox{\linewidth}{!}{
\begin{tabular}{|l|ccccc|c|}
\hline
Model Name & \multicolumn{5}{|c|}{ $R^2$ Score} & No. Parameters                         
\\ 
 & Subject 1 & Subject 2 & Subject 3 & Subject 4 & Average of All Subjects & \\ 
\hline

SVR (RBF)                          & 0.749 & 0.359 & 0.421 & 0.615 & 0.536 & N/A      \\
CNN                              & 0.814 & 0.725 & 0.797 & 0.682 & 0.755 & 337670  \\
LSTM (1 Layer)                    & 0.867 & 0.780 & 0.824 & 0.712 & 0.796 & 519212 \\
LSTM (2 Layer)                   & 0.876 & 0.782 & 0.827 & 0.737 & 0.806 & 840812 \\
LSTM (3 Layer)                   & 0.884 & 0.781 & 0.836 & 0.758 & \bf{0.815} & 1162412 \\
DPARS                          & 0.834 & 0.748 & 0.787 & 0.745 & 0.778 & \bf{6828} \\
DPARS (+entropy)    & 0.853 & 0.773 & 0.821 & 0.778 & 0.806 & \bf{6828} \\
\hline
\end{tabular}
}
\end{centering}
\\ 
\scriptsize{* Since the kernelized SVR is a non-parametric model, the number of parameters depends on the size of the training data \cite{russell2010artificial}.}
\vspace*{-2\baselineskip}
\end{table}
\subsection{Performance Evaluation and Comparison}
In this section, we evaluate the complexity and accuracy of the proposed DPARS model and compare it with other state-of-the-art EMG decoders.
The $R^2$ score and total number of parameters are used to measure the decoding accuracy and to estimate the model size, respectively. 
Table \ref{table: result} compares the performance of LSTM \cite{b15}, deep CNN  \cite{b11}, and support vector regression (SVR) \cite{pan2014continuous} with the proposed DPARS.

\begin{figure}
\centering
\includegraphics[width=.9\linewidth]{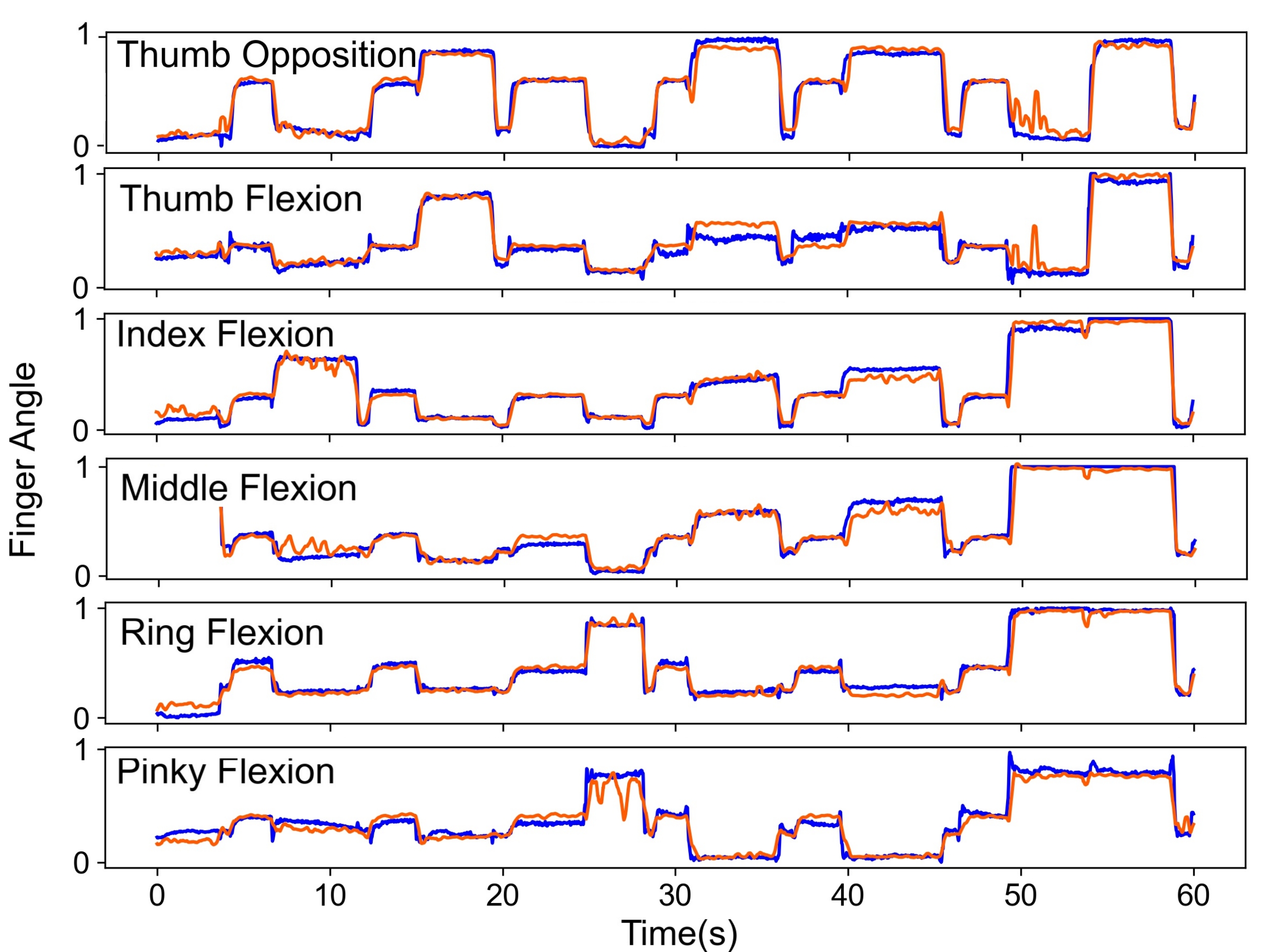}
\vspace*{-1\baselineskip}
\caption{\label{fig:output_compare} The real (blue line) and decoded (orange line) figure angles predicted by the proposed DPARS model.
\vspace*{-5\baselineskip}
}
\end{figure}

We found that SVR exhibits a low accuracy of 53.6\% on average, even with an RBF kernel.
CNN demonstrated a mean decoding accuracy of 75.5\%, while LSTM outperforms CNN and SVR with an accuracy of \textless81.5\%. 
These models are characterized by their large sizes, featuring several hundred thousand parameters.
Consequently, implementing such models in hardware would lead to large silicon area and high power consumption, making them impractical for next-generation lightweight RPHs.
\
Our novel attractor-based model decodes EMG signal with a mean accuracy of  80.6\% using a total of 6828 parameters. Figure \ref{fig:output_compare} illustrates the ground truth finger angles and the predicted output of the proposed decoder. This figure displays the decoder's proficiency in estimating finger movement angles across all angles, with the highest accuracy observed in the thumb opposition and ring flexion. These promising results show that DPARS achieves a comparable or superior performance in EMG decoding compared to  LSTM and CNN, respectively.
Furthermore, our proposed model, while \textgreater50$\times$ smaller than existing models, maintains competitive decoding performance and offers significant advantages in terms of reduced model complexity and energy efficiency.

\section{Conclusion}
This study proposed an innovative DPARS model for EMG decoding, 
employing a dual predictive strategy.
The model initially provides a coarse estimation based on attractors, capturing the underlying patterns in EMG signals, and then refines this estimation using a lightweight regressor. 
Owing to the extensive dimensionality reduction, weight sharing, and dual predictive strategy, DPARS exhibits a remarkable reduction in model size, more compact than LSTM and CNN models by a factor of 50 to 120, respectively.
Furthermore, the decoder network only needs to compute the attractors' likelihood (i.e., not for all states), resulting in more than a fourfold reduction in the number of MAC operations.
We proved  DPARS is accurate, compact, and low-complexity, making it well-suited 
for lightweight and accessible AI-enabled hand prostheses
that can significantly improve the quality of life of amputees.

\vspace*{-1\baselineskip}

\end{document}